\documentclass[]{emulateapj}

\shorttitle{Coma galaxy orientations}
\shortauthors{Torlina et al.}

\begin{document}


\title{Galaxy orientations in the Coma cluster}


\author{Lisa Torlina}
\affil{School of Physics, University of Sydney, NSW 2006, Australia}
       
\author{Roberto De Propris}
\affil{Cerro Tololo Inter-American Observatory, 
Casilla 603, La Serena, Chile}
       
\author{Michael J. West}
\affil{Gemini Observatory, Casilla 603, La Serena, Chile}


\begin{abstract}
We have examined the orientations of early-type galaxies in the Coma
cluster to see whether the well-established tendency for brightest cluster
galaxies to share the same major axis orientation as their host cluster also
extends to the rest of the galaxy population.  We find no evidence of any
preferential orientations of galaxies within Coma or its surroundings. 
The implications of this result for theories of the formation of clusters 
and galaxies (particularly the first-ranked members) are discussed.

\end{abstract}




\section{Introduction}

It has long been known that the major axis of central cluster galaxies 
tends to be aligned with the major axis of the cluster galaxy distribution
and to point towards similar structures on scales of typically a few
tens of Mpc (e.g., \citealt{binggeli82,west89}; Trevese, Cirimele \&
Flin 1992; Fuller, West \& Bridges 1999, \citealt{plionis03} and references
therein). The most likely mechanism for the origin of this `alignment effect'
invokes {\it collimated infall} of galaxies along the filaments that make 
up the large scale structure in which clusters are embedded and from which 
clusters grow \citep{west94,dubinski98}, suggesting a connection between the 
growth of the clusters and the formation of their massive dominant galaxies
(e.g., \citealt{faltenbacher05}). 

It is unclear whether fainter cluster galaxies should show the alignments
that are expected (and observed) for the first-ranked members. Clusters and
galaxies are believed to form through a process of hierarchical accretion.
In these models, predictions for isotropy (or deviations) of galaxy position
angles are not well defined (West, Villumsen \& Dekel 1991) but the consensus
seems to be that any primordial alignments would be quickly erased by
dynamical interactions \citep{coutts96,plionis03}. Conversely, it is
expected that groups and clusters show preferential alignments in the
direction of the last major accretion episode \citep{faltenbacher05}.
Searches for alignments among the fainter cluster galaxies have generally
returned ambiguous results, with some positive detections \citep{plionis03,
aryal06} as well as null results (\citealt{strazzullo05}; Aryal, Kandel 
\& Sauer 2006). However, it is possible that the presence of anisotropy in
galaxy position angles can only be detected in less evolved clusters 
\citep{plionis03}, while it is also conceivable that only near neighbor
galaxies are aligned with each other because they exert reciprocal tidal
torques \citep{kitzbichler03}, or that radial alignments are induced by
the cluster tidal field \citep{pereira05}.

With this in mind, we re-examine the issue of isotropy in galaxy position
angles in the Coma cluster. We use a unique multicolor optical-infrared 
CCD mosaic of the inner 750 $h^{-1}$ Mpc of this cluster and, within this
region, a complete (magnitude-limited to $H < 14.5$) sample of 111 spectroscopic
members plus 77 additional fainter members. Most previous work has
relied on photographic plates and identified cluster members statistically
or via the red sequence. The results we present in the following are most
consistent with an isotropic distribution for galaxy position angles, other
than for the two brighter members. We then explore the implications of these
findings for the formation of clusters and their members, particularly as
regards the brighter central cluster galaxies.

\section{Data analysis}

We use the $UBVRIzJHK$ images of the core of the Coma cluster and the
catalog published by \cite{eisenhardt07}, which includes photometry and 
a compilation of spectroscopy from the literature. The spectroscopic sample 
of Coma galaxies is complete to $H=14.5$ and less so for fainter magnitudes.
  
Galaxy major axis position angles and ellipticities obtained independently from 
$U$, $V$ and $K$ images were compared and found to be similar; consequently 
only the $V$-band images were used for further analysis.
We verified 
the SExtractor position angles by isophotal modelling of the galaxies using IRAF ELLIPSE 
and also by comparing with 
the isophotal position angles from the SDSS \citep{york00,stoughton02}. 
The values produced by SExtractor are in good agreement with the measurements
from IRAF ELLIPSE and the SDSS position angles, except for a few objects
with small ellipticities.

We tested the observed distribution of position angles against a uniform
distribution using the Kuiper test. Figure 1 shows the results of this
comparison: we find no significant deviation from isotropy. This holds if
we limit our sample to galaxies with larger ellipticity or to the brighter
members (e.g., \citealt{west98}).  Subsamples of galaxies with ellipticities 
$e > 0.1, 0.2$ and $0.3$ were examined, containing 128, 94 and 61 
galaxies, respectively, but in all cases no preferred orientations were found.
Likewise culling the sample of galaxies to those brighter than $H=11, 12$ and 
$13$ (10, 33 and 58 galaxies, respectively) also yielded a null detection 
of position angle anisotropies.

\begin{figure}
\epsscale{0.90}
\plotone{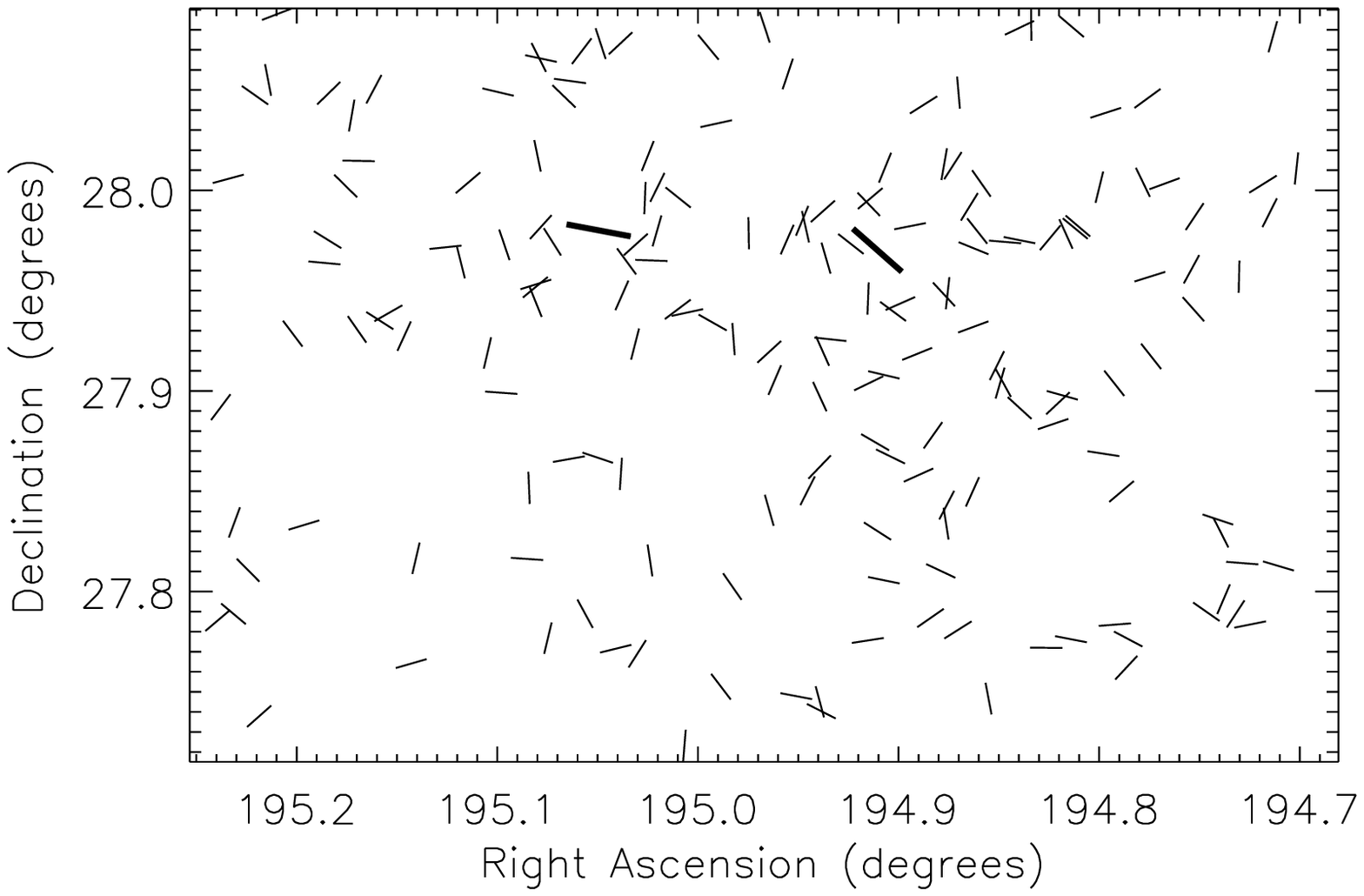}\\
\plotone{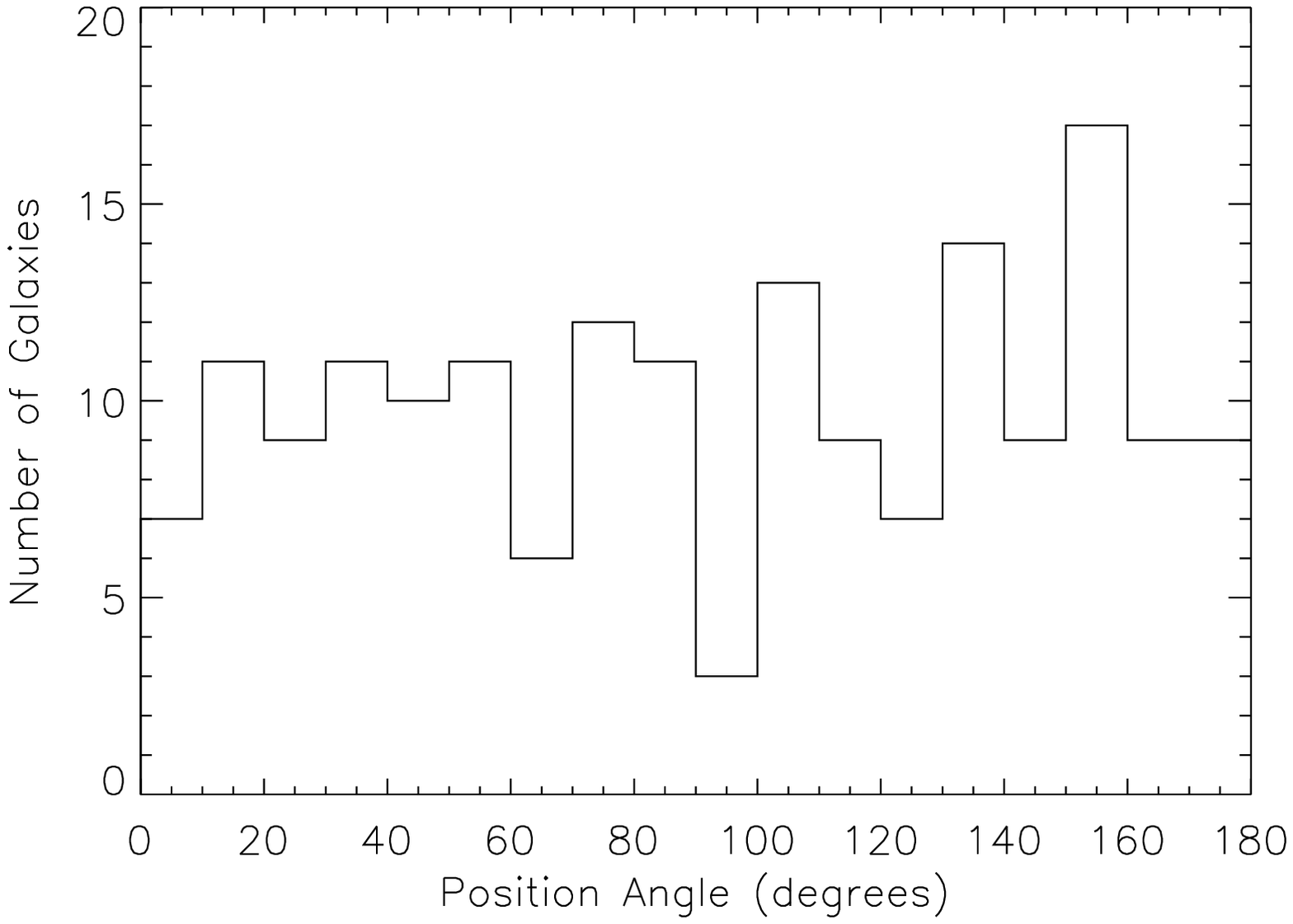}\\
\plotone{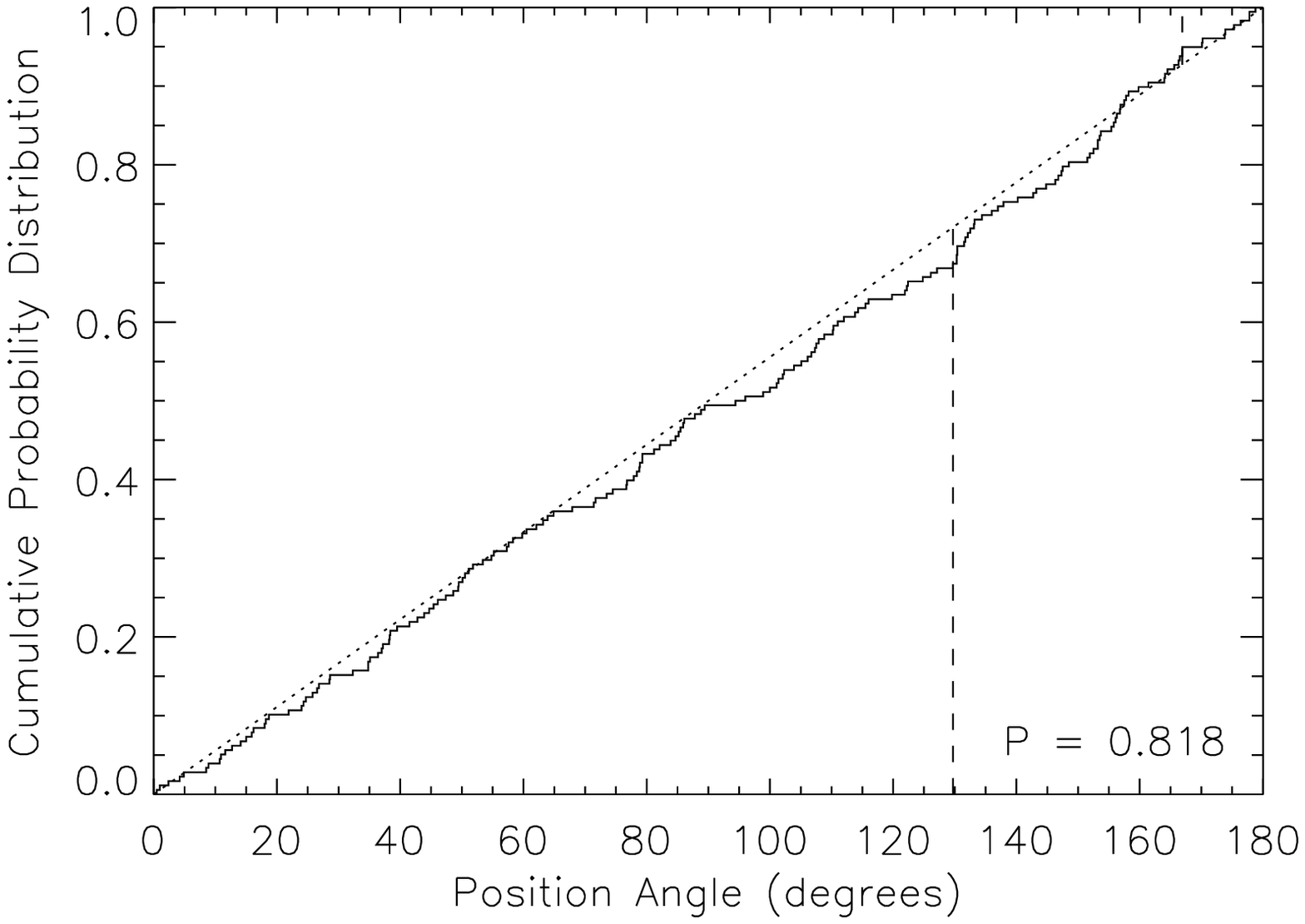}
\caption{Top Panel: Positions of galaxies in the Coma field; the segments 
represent individual galaxies and are oriented according to its position
angle. The two thick segments identify the two brightest cluster members.
Middle Panel: Histogram of the position angles for Coma cluster members.
Bottom Panel: Probability distribution and comparison with a uniform distribution;
the number at bottom right indicates the likelihood that the observed distribution
is uniform: in this case 82\%}
\end{figure}

While the above constitutes strong evidence that there are no large-scale
galaxy alignments (other than for the brightest cluster galaxies), we now
consider whether there are alignments between nearest neighbors (in projection)
due to reciprocal tidal torques \citep{kitzbichler03}, or due to the 
presence of substructure whose members are aligned with each other \citep{plionis02}. The
result of this test is shown in Figure 2, where we find no significant
deviation from isotropy. Neither can we confirm the detection of radial
galaxy alignments by \cite{pereira05}. The above results are perhaps understandable
if we consider that there is no reason to assume that cluster members that 
are close in projection are actually physically close, because clusters
have intrinsic depth and an extended radial velocity distribution. Therefore,
any true alignment signal between neighbors will be diluted by the overwhelming
number of galaxy `asterisms'.

\begin{figure}
\plotone{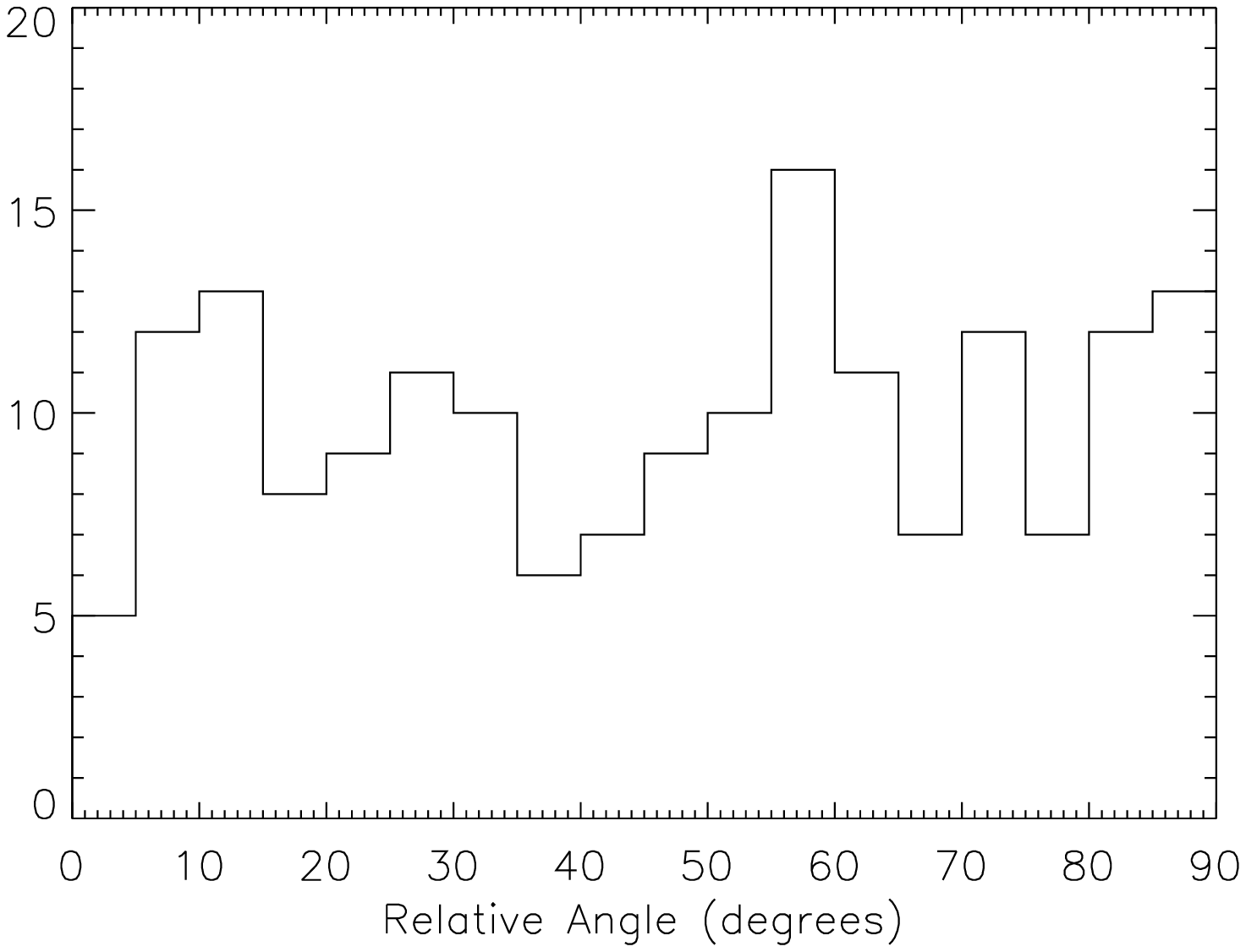}\\
\plotone{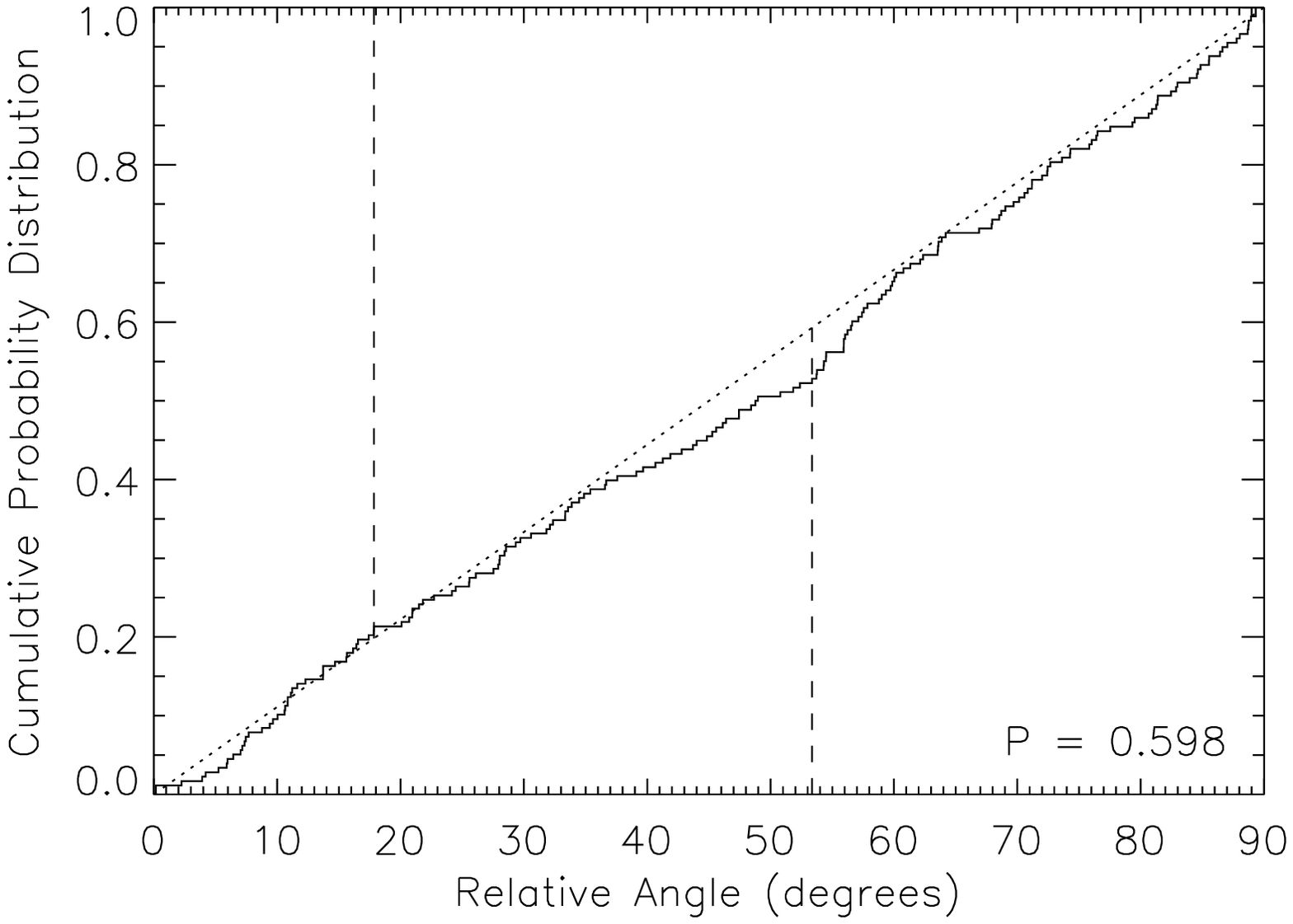}
\caption{Histogram of position angles and probability distribution
for the alignments of nearest neighbor galaxies}
\end{figure}

\section{Discussion}

The results of the analysis presented above are consistent with random orientations
of the position angles of Coma galaxies, with the possible exception of the 
two brightest cluster members, NGC 4889 and NGC 4874, whose major axis
position angles are quite similar to that of the cluster's galaxy and
X-ray gas distributions \citep{west00}. The line connecting the two galaxies
is also aligned with the galaxy and X-ray gas distributions, as also found
by \cite{trevese92}. While this is consistent with the null results found
for some clusters, there are other objects where significant alignments are
detected. Possibly the clearest case is Abell 521 \citep{plionis03}, which
shows detectable alignments out to 5 $h^{-1}$ Mpc. This cluster is known
to be dynamically young and to show considerable substructure \citep{ferrari06}.
Coma, on the other hand, is not a fully relaxed cluster and shows substructure
in the X-ray gas and in the position-velocity distribution of its members 
(e.g., \citealt{adami05}) but it is clearly (comparatively) more evolved than 
either A521 or Virgo, where also there is evidence of a non-isotropic distribution 
of galaxy alignments \citep{west00}. As suggested by \cite{plionis03}, the lack 
of alignments in Coma may be due to its more advanced state of dynamical evolution
with respect to other clusters (see, e.g, \citealt{aryal06}). 

\begin{figure}
\plotone{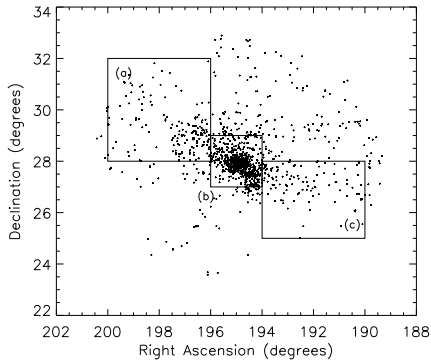}
\caption{Coma and its surroundings: the boxes mark the extended regions connecting
Coma to its substructure drawn from the SDSS data}
\end{figure}

In order to test this, we considered galaxy alignments in the filaments
that connect Coma to the large scale structure towards A1367 and A2197/A2199
(Figure 3). We retrieved these data from the SDSS and tested for position 
angle anisotropy.  All galaxies with available data were used, irrespective 
of their magnitudes.  Because complete SDSS spectroscopy is not available for this region,
redshifts were taken from the NASA Extragalactic Database and the sample of 
galaxies restricted to those with velocities in the range 4000 - 10,000 km/s 
expected for Coma and its environs.

\begin{figure*}
\begin{tabular}{ccc}
\includegraphics[width=2.1in]{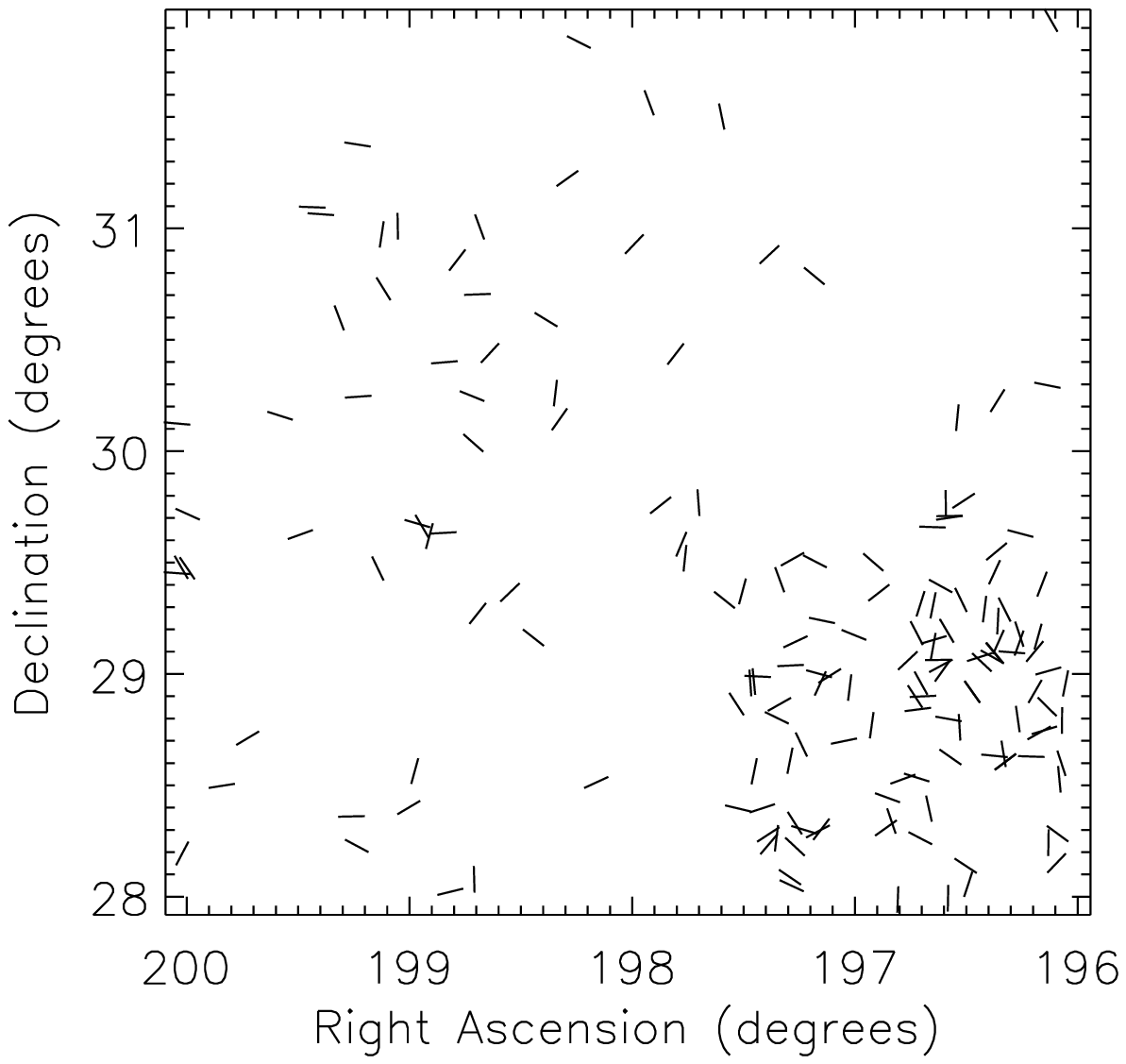} & \includegraphics[width=2.1in]{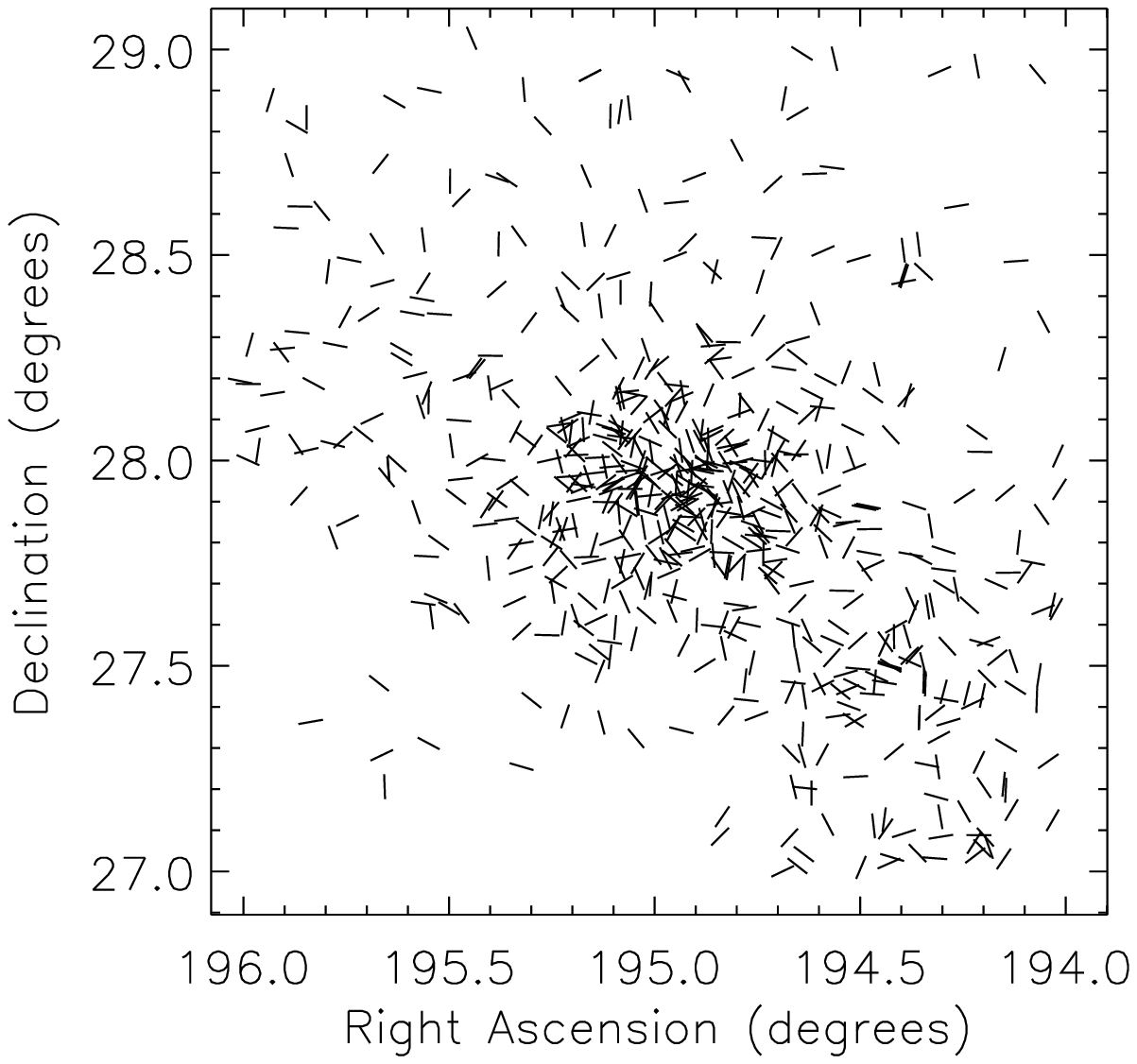} &
\includegraphics[width=2.1in]{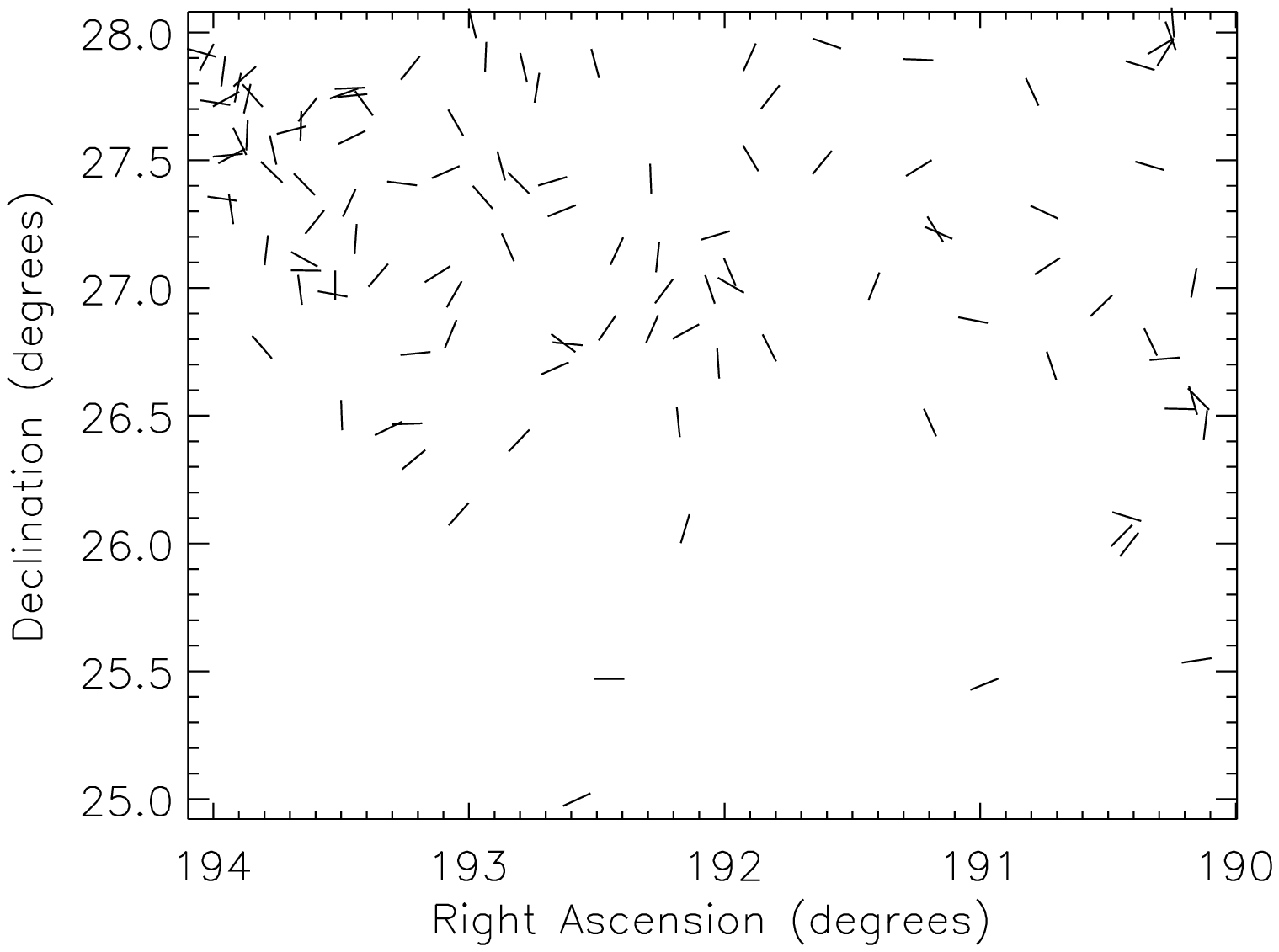} \\
\includegraphics[width=2.1in]{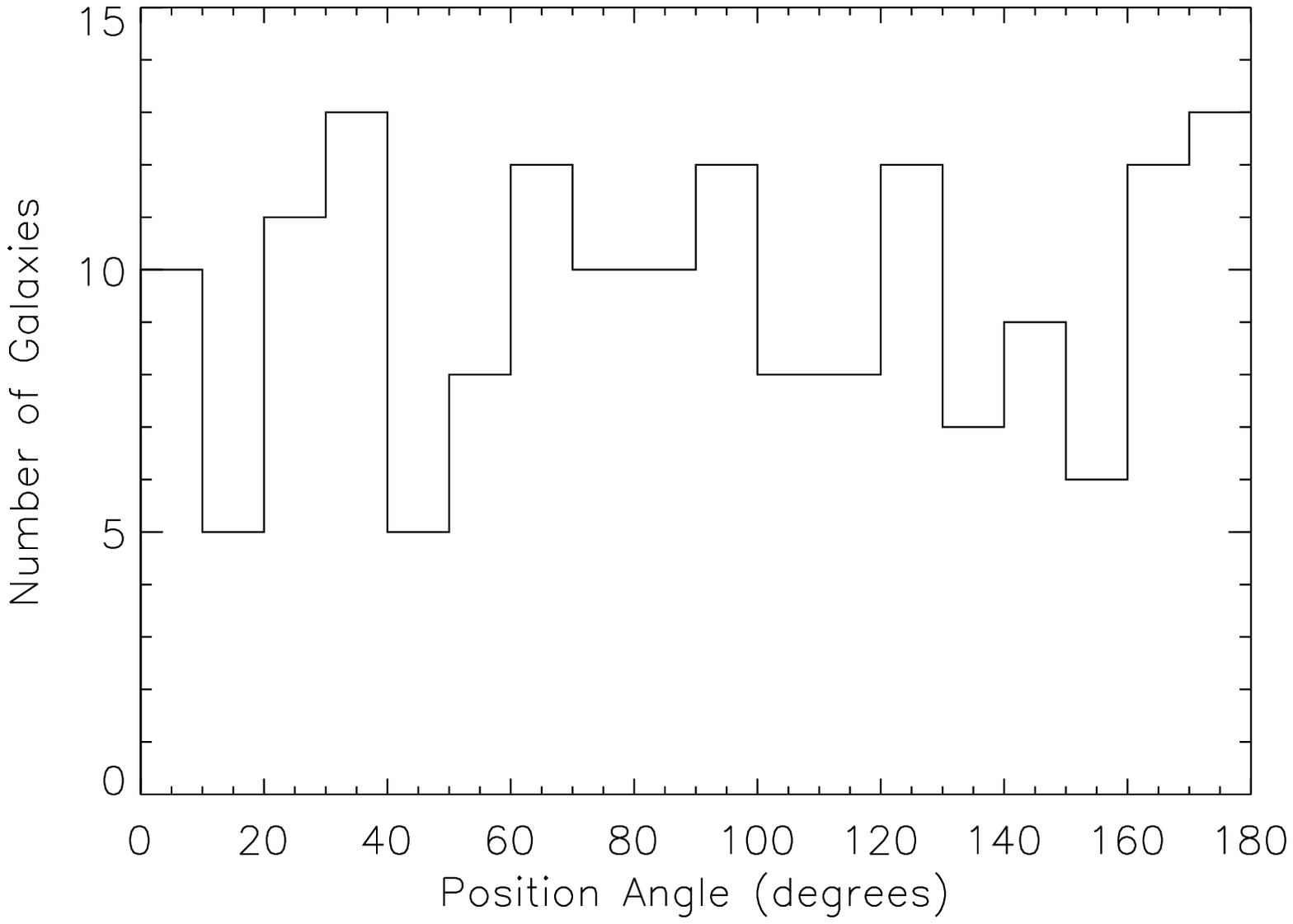} & \includegraphics[width=2.1in]{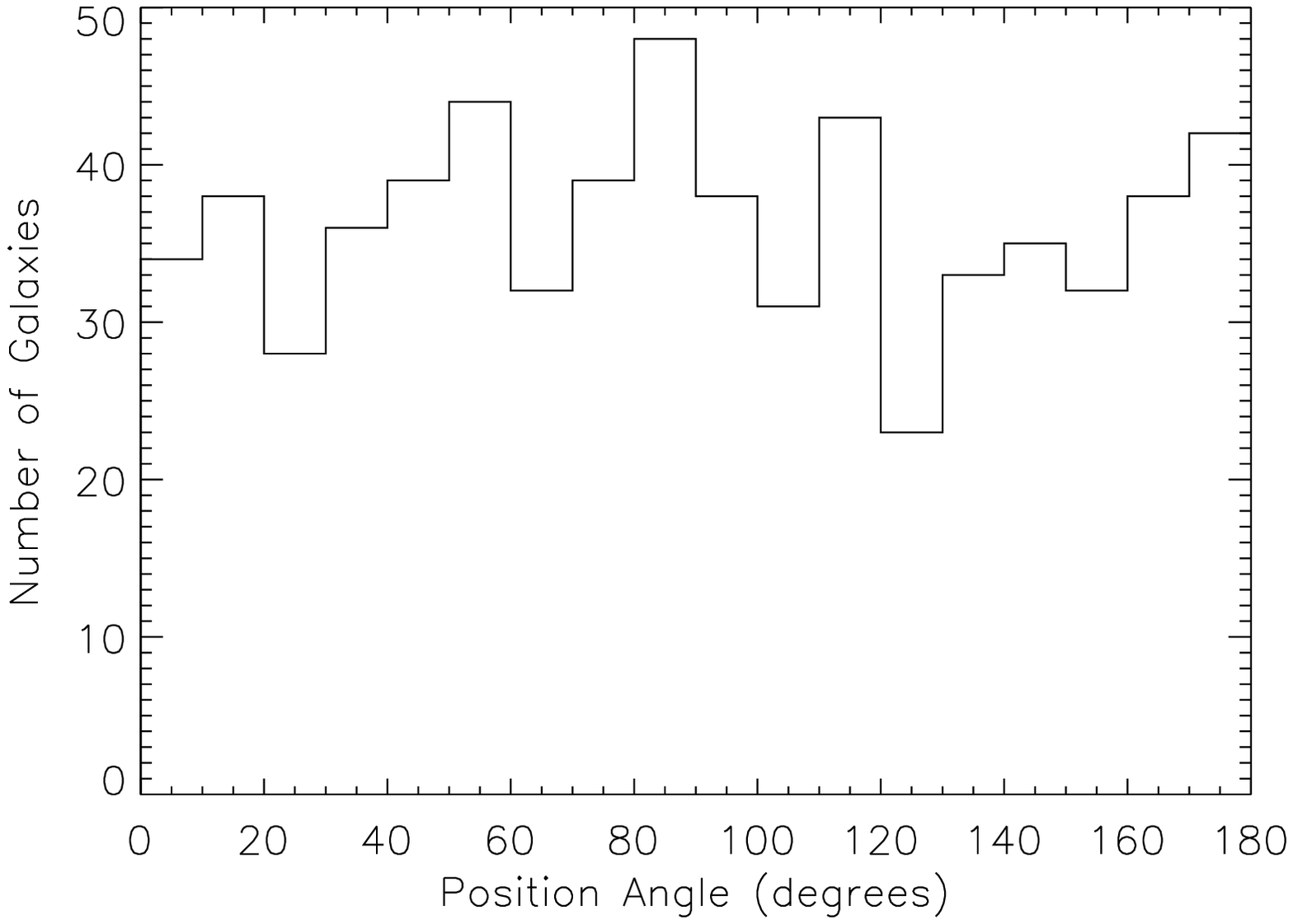} &
\includegraphics[width=2.1in]{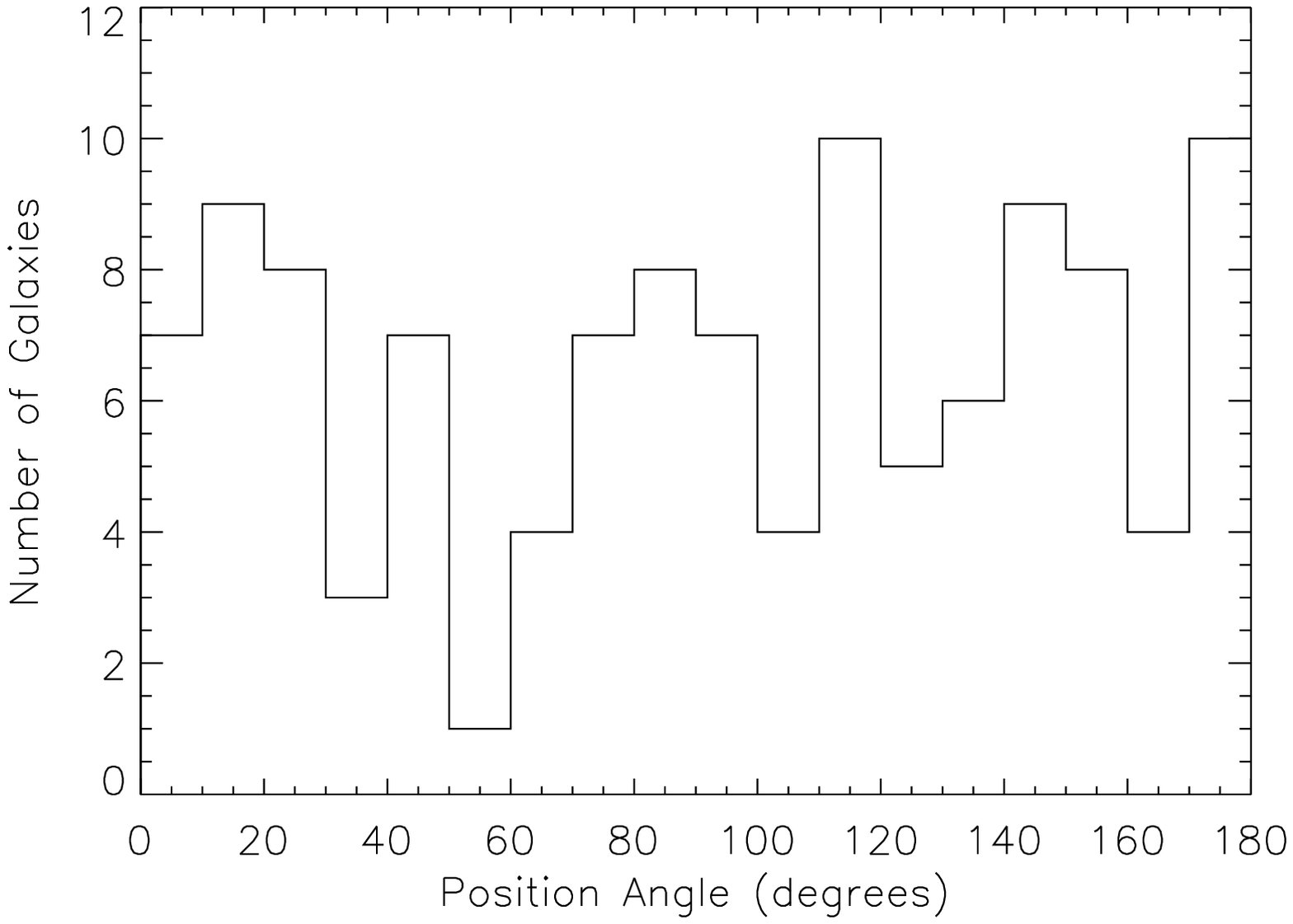} \\
\includegraphics[width=2.1in]{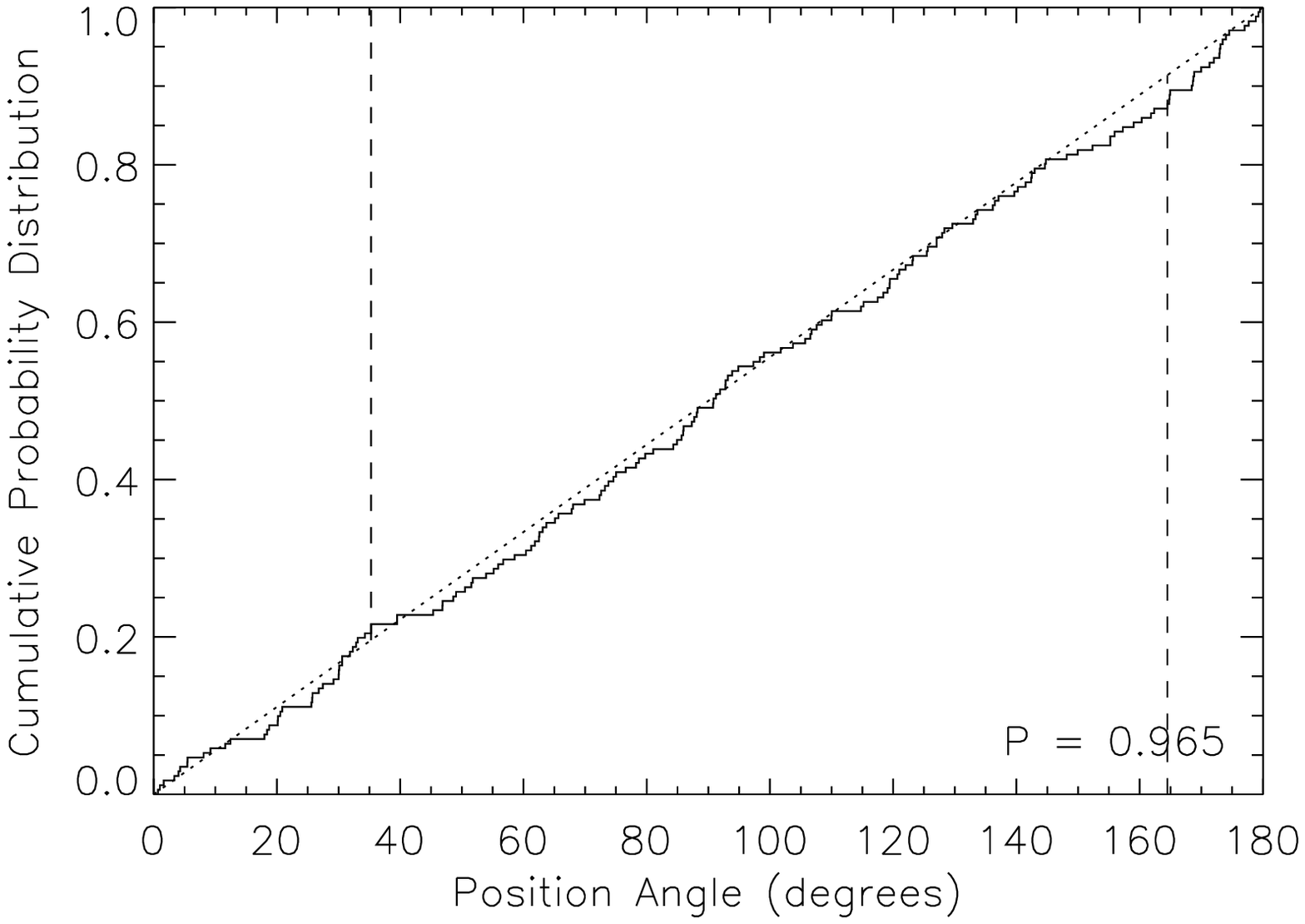} & \includegraphics[width=2.1in]{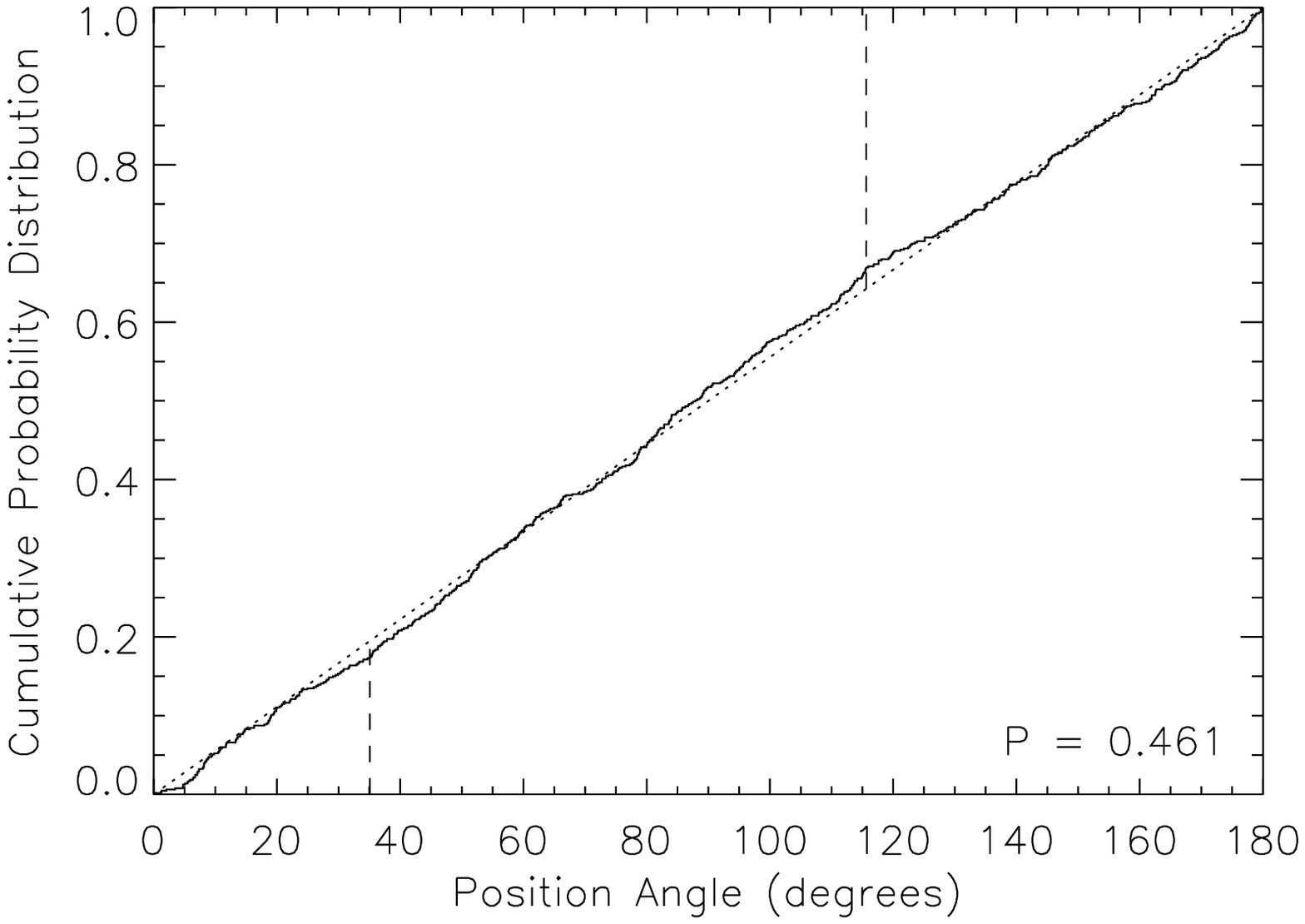} &
\includegraphics[width=2.1in]{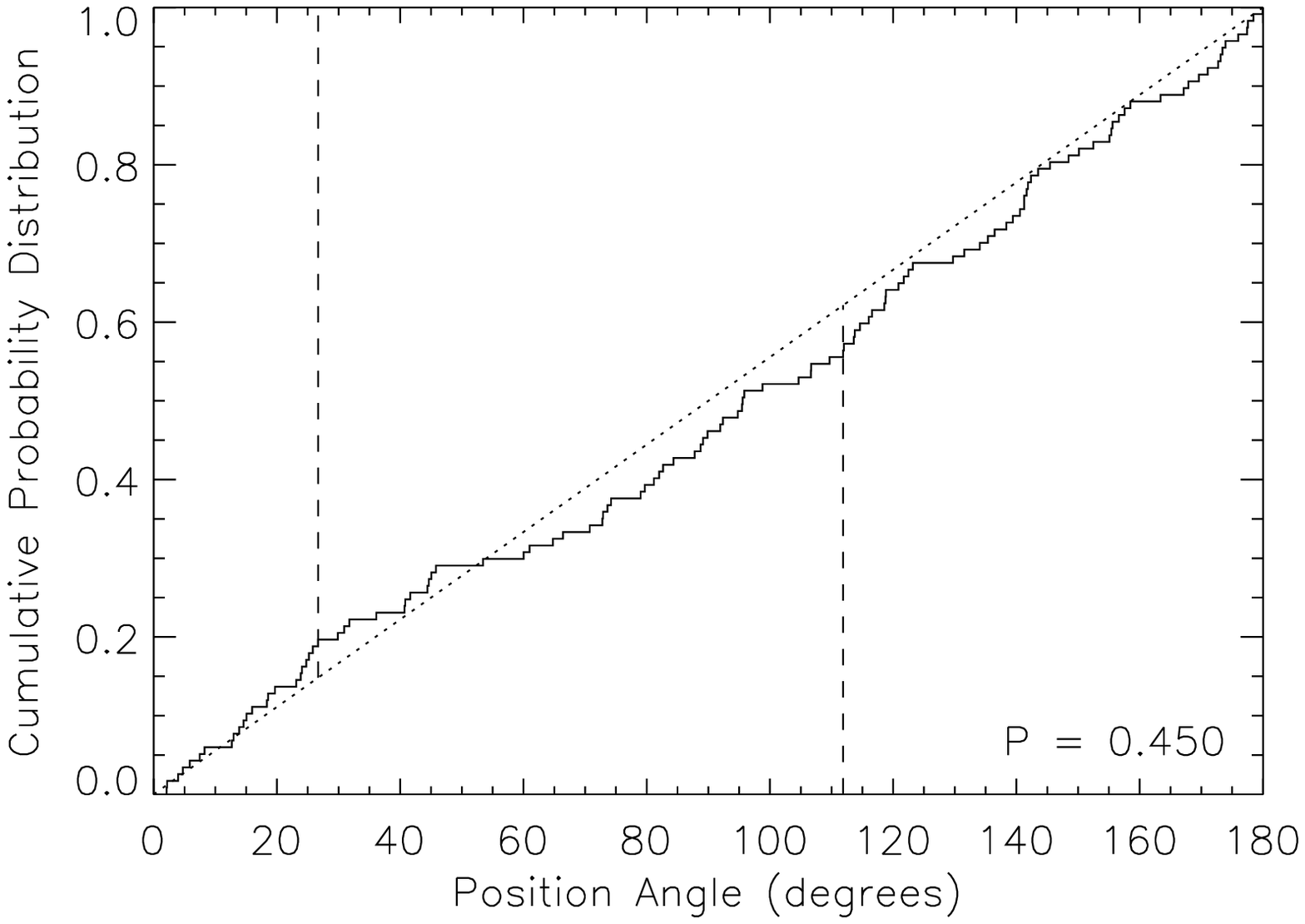} \\
\end{tabular}
\caption{Top panels: the positions of the galaxies in the boxes (marked a,b,c)
of Figure 3 and their position angles, marked as segments, as for Figure 1.
Middle panels: Histograms of the position angles. Bottom panels: probability
distributions compared to a uniform distribution.}
\end{figure*}

  Here too, however, we find no evidence for any significant 
deviations from isotropy for filament galaxies (Figure 4). This suggests either 
that these galaxies formed without any preferred orientations or that any 
alignments which might have existed are quickly erased even in low density 
regions. In this context, it is not surprising that alignments can be detected 
more clearly in A521 and Virgo, two clusters that appear to be have recently 
formed from the coalescence of numerous small groups.

The main inference we can draw from this study is that the first-ranked
cluster galaxies are special in yet one more way. Unlike all other cluster
galaxies they appear to have undergone a peculiar formation mechanism,
related to collimated infall and the growth of the cluster from the
surrounding large scale structure, and resulting in the observed alignment
effects. All other galaxies appear to have had a more random merger history.
While fainter galaxies may have been accreted to the cluster at later epochs,
thus explaining the lack of alignments, the brightest cluster member may
have been the kernel around which clusters originally formed.

\section*{Acknowledgements}
L. Torlina was supported by an Australian Gemini Undergraduate Summer Studentship,
 funded by the Australian Research Council.  She thanks 
Gemini Observatory for its hospitality during her stay in Chile.  We thank the referee, 
Manolis Plionis, for his very quick 
review of this paper and for helpful comments that improved its clarity.

\setlength{\bibhang}{2.0em}

\clearpage

\end{document}